\documentclass[10pt,a4paper]{article}
\usepackage[dvips]{color}
\usepackage{epsfig}
\usepackage{amsmath}
\usepackage{graphicx}
\usepackage{authblk}
\usepackage{amssymb,amsmath}
\usepackage{amsmath}

\begin{document}
\title{\bf Cosmological model involving an interacting van der Waals fluid}
\author{E. Elizalde$^{a,b,c}$\thanks{E-mail: elizalde@ieec.uab.es}, {M. Khurshudyan$^{b,d,e}$\thanks{Email:khurshudyan@yandex.ru, khurshudyan@tusur.ru}}\\
$^a${\small {\em Consejo Superior de Investigaciones Cient\'{\i}ficas, ICE/CSIC-IEEC,
Campus UAB, Carrer de Can Magrans s/n, 08193 Bellaterra (Barcelona) Spain}}\\
$^{b}${\small {\em International Laboratory for Theoretical Cosmology, Tomsk State University of Control Systems and Radioelectronics (TUSUR), 634050 Tomsk, Russia}}\\
$^{c}${\small {\em  Kobayashi-Maskawa Institute, Nagoya, Japan}}\\
$^{d}${\small {\em Research Division,Tomsk State Pedagogical University, 634061 Tomsk, Russia}}\\
$^{e}${\small {\em Institute of Physics, University of Zielona Gora, Prof. Z. Szafrana 4a, 65-516 Zielona Gora, Poland}}\\
}

\maketitle

\begin{abstract}
A model for the late-time accelerated expansion of the universe is considered where a van der Waals fluid interacting with matter plays the role of dark energy. The transition towards this phase in the cosmic evolution history is discussed in detail and, moreover, a complete classification of the future finite-time singularities is obtained for six different possible forms of the non-gravitational interaction between dark energy (the van der Waals fluid) and dark matter.
This study shows, in particular, that a universe with a non-interacting three-parameter van der Waals fluid can evolve into a universe characterized by a Type IV~(Generalized Sudden) Singularity. On the other hand, for certain values of the parameters, exit from the accelerated expanding phase is possible in the near future, what means that the expansion of the universe in the future could become decelerated. On the other hand, our study shows that space can be divided into different regions. For some of them, in particular,  the non-gravitational interactions $Q = 3 H b \rho_{de}$, $Q = 3 H b \rho_{dm}$ and $Q = 3 H b (\rho_{de} + \rho_{de})$ may completely suppress future finite-time singularity formation, for sufficiently high values of $b$. On the other hand, for some other regions of the parameter space, the mentioned interactions would not affect the singularity type, namely the Type IV singularity generated in the case of the non-interacting model would be preserved. A similar conclusion has been archived for the cases of $Q = 3 b H \rho_{de}\rho_{dm}/(\rho_{de}+\rho_{dm})$, $Q = 3 b H \rho_{dm}^{2}/(\rho_{de}+\rho_{dm})$ and $Q = 3 b H \rho_{de}^{2}/(\rho_{de}+\rho_{dm})$ non-gravitational interactions, with only one difference: the $Q = 3 b H \rho_{dm}^{2}/(\rho_{de}+\rho_{dm})$ interaction will change the Type IV singularity of the non-interacting model into a Type II~(The Sudden) singularity.
\end{abstract}

\section{Introduction}

In this paper we address the key problem of explaining the acceleration of the universe expansion under the general framework of general relativity as the theory describing the background dynamics. Dark energy will be for us an extra component, and not just fluctuations of the quantum vacuum (which are also problematic, on their own side). Ongoing intense theoretical research and ever more accurate astronomical observations have not yet been able to focuss on a generally accepted solution of this very important problem. Of course, a perfectly acceptable alternative is to resort to a modification of general relativity, playing a similar role to dark energy. The questions of how to model dark energy or how to precisely modify general relativity in order to obtain a picture both consistent with recent observational data and coming from some fundamental solution remains still open.

This situation gives a rise to the appearance of many different candidates for dark energy models and modified theories of gravity~\cite{Od1}~(and references therein). Besides their role as cosmological models to explain the different stages of the universe evolution, they have also been checked for more detailed astrophysical applications~\cite{Od3}~(to mention a few). This paper  will be devoted to a model of dark energy as a dark fluid, known in the literature as a van der Waals fluid model, and we will start with a very brief presentation of the main aspects and developments of this subject. This will be complemented with a basic list of  references on modified theories of gravity, together with their cosmological and astrophysical applications, which we will divide into three main groups.
In the first one, we gather scalar field dark energy models like quintessence, phantom, the quintom model, and others. In the second group we include dark energy models which are described by the energy density. Some of the examples here are holographic dark energy models, including a more general class  known as Nojiri-Odintsov holographic dark energy models, and varying and usual ghost dark energy models. Finally, the Chaplygin, polytropic and van der Waals dark energy models are some of the examples included in the third group~(see, for instance, \cite{M1} and \cite{M2} and the references therein).

There are numerous studies dealing with one or the other of these models so that, to save space, we refer the readers to the corresponding references and concentrate directly in the case of theories of gravity involving a van der Waals fluid as dark energy. The study of this class of models, where the van der Waals fluid plays either the role of the whole energy source, or of just a part of it, motivate the present work. In particular, one of the first analysis involving a free parameter van der Waals fluid was  discussed in Ref.~\cite{W1}, where the universe was modeled by a binary mixture consisting of a van der Waals fluid and another dark energy component, namely either quintessence or a Chaplygin gas. Moreover, it was taken there into account the irreversible processes occurring in the form of an energy transfer between the van der Waals fluid and the gravitational field. Among other results, this study showed that the model could simulate:
\begin{enumerate}
\item an inflationary period where the cosmic acceleration grows exponentially and the van der Waals fluid behaves like an inflaton,
\item an accelerated period where the acceleration is positive but it decreases and tends to zero whereas the energy density of the van der Waals fluid decays,
\item a decelerated period which corresponds to a matter dominated epoch with a non-negative pressure,
\item a present  stage of accelerated expansion, where the dark energy density outweighs the energy density of the van der Waals fluid.
\end{enumerate}
In Ref.~\cite{W2} the viability of three different parameterizations of the van der Waals fluid as a model of inflation in the early universe (containing one, two, and three free parameters, respectively)  was considered. This study shows that a de Sitter-like expansion in the early universe will be observed in the models discussed. Detailed analysis of Ref.~\cite{W2} yields a clearly understanding of when a given model can be ruled out, in particular, it shows that two of the three models considered violate some observational constraints, and that the third model (the one with three free parameters) is highly fine-tuned. On the other hand, in Ref.~\cite{W3} theoretical fluid models obeying van der Waals equation of state for inflation in the presence of viscosity have been discussed. The validity of the models there were tested via comparison of the most fundamental inflationary parameters, namely the spectral index and the tensor-to-scalar ratio, with the latest results coming from accurate data of the Planck satellite. After imposing some restrictions on the parameters, one can check that the authors did obtain good agreement with the astronomical data. The study of Ref.~\cite{W3} clearly demonstrates that the inclusion of viscosity (usually neglected for the inflationary epoch) in a van der Waals fluid model definitely enhances the agreement with the results of astronomical observations, in a smooth and natural way. Some other aspects obtained during the study of cosmological models with a van der Waals gas can be found in Ref.~\cite{W4}.

After doing some mathematics, we reach the conclusion that the universe's Hubble parameter and the one-parameter van der Waals component, in terms of the deceleration parameter $q$ and the model parameter $\omega$,can be expressed as a solution of the following equation
\begin{equation}\label{eq:H12}
\frac{1}{2} H^2 \left(\frac{3 \omega }{3 H^2-1}+9 H^2+2 q-1\right) = 0.
\end{equation}
Interesting solutions representing an expanding universe in this case are
\begin{equation}
H_{1,2} = \frac{1}{3} \sqrt{2 - q \pm \sqrt{q^2+2 q-9 \omega +1}}.
\end{equation}
To obtain Eq.~(\ref{eq:H12}) one takes into account that $\dot{H} = -H^{2}(1+q)$. On the other hand, with $\ddot{H} = H^{3} (j + 3q + 2)$ and $\dot{H} = -H^{2}(1+q)$, one gets a system of  equations allowing to express the Hubble parameter $H$ and the model parameter $\omega$ in terms of $q$ and $j$, where $j$ is the jerk parameter defined from (with $n = 3$)
\begin{equation}
C_{n} = \frac{1}{a}\frac{d^{n}a}{dt^{n}}H^{-n},
\end{equation}
 $a$ being the scale factor.
A significant number of papers, some of which have been already  mentioned above, directly demonstrate the viability of a van der Waals fluid. However,  to our knowledge a classification of future finite-time singularities for the class of cosmological models with a van der Waals fluid is still lacking. The main purpose of the present paper is to fill in part  this gap by providing such a classification for a three-parameter family of models with a van der Waals dark fluid, namely
\begin{equation}\label{eq:Waals}
P_{de} = \frac{\omega \rho_{de}}{1-\alpha \rho_{de}} - \beta \rho^{2}_{de},
\end{equation}
being $\omega$, $\alpha$ and $\beta$ the free parameters, while $\rho_{de}$ is the energy density of the dark fluid, a part of the whole energy content of the present and recent universe. In general, it is  believed that a classification of future singularities of a cosmological model will be useful for better understanding the physics of the universe. Also, this can be used in order to understand if the recent observational data could eventually provide a unique connection between physics of the early and future universe, based on appropriate theories developed for our present universe. The answer will depend in part on the statistical error of the particular data and on the possible tensions that may arise among the  different observational data sets. In what follows, all types of future finite-time singularities arising in this family of models, together with some further discussion on their evolution and some ensuing practical results will be presented.

To start, we know that in a universe with phantom fields, with interacting dark matter/dark energy or described by theories of modified gravity, the following future finite-time singularities arise~\cite{S1}
\begin{itemize}
\item Type I singularity~("The BIg Rip Singularity"): If the singularity occurs at time  $t = t_{s}$, then the scale factor $a$, the effective energy density $\rho_{eff}$ and the pressure $P_{eff}$, diverge as $t \to t_{s}$; that is, $a \to \infty$, $\rho_{eff} \to \infty$ and $|P_{eff}| \to \infty$. This case yields incomplete null and time-likegeodesics,
\item Type II singularity~(\^{a}€{\oe}The Sudden Singularity): In this case, the scale factor $a$ and the total effective energy density $\rho_{eff}$ are finite, but the effective pressure $P_{eff}$ diverges as $t \to t_{s}$; that is, $a \to  a_{s}<\infty$, $\rho_{eff} \to \rho_{s}<\infty$ and $|P_{eff}| \to \infty$. In this case the geodesics are complete and observers are not necessarily crushed (weak singularity),
\item Type III singularity~("The Big Freeze Singularity"): In this case, only the scale factor is finite, and the effective pressure and effective density diverge at $t \to t_{s}$; that is, $a \to  a_{s}<\infty$, $\rho_{eff} \to \infty$ and $|P_{eff}| \to \infty$. These can be either weak or strong singularity, which are geodesically complete solutions,
\item Type IV singularity~("Generalized Sudden Singularity"): In this case, the scale factor, the effective pressure and the effective density are finite at $t \to t_{s}$. On the other hand, the Hubble rate and its first derivative are also finite, but the higher derivatives of the Hubble rate diverge at $t \to t_{s}$. In this case there occur weak singularities and the geodesics are complete,
\end{itemize}
Singularities can appear as a consequence of the violation of the null energy condition. Recently, the quantitative implications of finite-time singularities on the amplitude of the primordial gravitational waves have been considered in Ref.~\cite{S2}. In particular it has been found there that in the case of F(R) theories, a Type IV finite-time singularity has no effect on the gravitational waves, i.e. the amplitude of the gravitational wave modes are unaffected. Moreover, the universe smoothly goes through the singularity and the Type IV singularities can generate a graceful exit from inflation, altering its own dynamics. For the results of studies in the case of a Type II finite-time singularity we refer the reader to Ref.~\cite{S2}. Some developments about finite-time singularities in F(R) gravity can be found in~\cite{S2_1}. On the other hand, in Ref.~\cite{S3}, using SN Ia, BAO and H(z) data, a general lower bound for the time of a potential future singularity has been obtained. Among other results of this study, one sees that a potential future singularity cannot be closer to the present time than $\sim 0.2t_{0}$, which roughly corresponds to $2.8$ Gyr, and a consistent lower bound yields around $1.2 - 1.5t_{0}$ (for certain models of parametrization). These conclusions could be extended; in particular, the authors of Ref.~\cite{S2} already indicated that it would be interesting to consider $f(G)$ theories of gravity. This consideration is mandatory for a better understanding of the issue, but at the same time it is necessary to consider, for instance, F(R,T), F(G,T), and F(R,G) modified theories of gravity, to draw a further comprehensive and self-consistent picture of this topic. This study can be useful also for distinguishing the already mentioned modified theories of gravity from each other and, probably, it will allow to highlight some hidden features. On the other hand, the study of Ref.~\cite{S3} could be extended in two direction. The first and obvious one is a consideration of new parameterizations in light of new developments in theoretical cosmology. While the second path of the extension could be related to the inclusion of strong gravitational lensing data that are being intensively used in the recent literature.

Finite-time singularities could be formed in interacting dark energy models, as well. There are several reasons for introducing non-gravitational interactions between dark energy and dark matter. One of them is the dependence of the solution of the cosmological coincidence problem on the form and existence of non-gravitational interactions (see, for instance, \cite{M1,M2}). The stabilization of the early energy source would be the second argument supporting the existence of a non-gravitational interactions, as well. It is also important to reconstruct the form of the non-gravitational interactions directly from the observational data, instead of using phenomenological assumptions, as is done in various works. In this case, only the form of the dark energy EoS will be enough and it is important to have a well implemented, model-independent data-smoothing tool, like the Gaussian Processes \cite{GP}.

As is well known, non-gravitational interactions affect the background dynamics and, in principle, the main characteristics of finite-time singularities could be altered, too. Therefore it is very important to have a classification of singularities for a cosmological model with an interacting dark energy, because an imprint of the non-gravitational interaction in the mechanism of singularity formation would be useful also for understanding the nature of the interaction itself. But this is quite important, since nowadays the non-gravitational interaction is understood as an energy transfer between the energy sources, i.e., an energy transfer either from dark energy to dark matter, or from dark matter to dark energy. Due to the existing symmetries and well known tensions, the question related to the non-gravitational interactions remains unanswered to a large extent.

In this paper we will study how different forms of the non-gravitational interaction affect the type and characteristics of the finite-time singularities. In particular, we aim at showing that the Type IV~(Generalized Sudden) singularity formed in the case of a non-interacting three-parameter van der Waals dark fluid could be changed into a Type II~(The Sudden) singularity if the non-gravitational interaction is of the form
\begin{equation}\label{eq:QM23}
Q = 3 b H \frac{\rho_{dm}^{2}}{(\rho_{de}+\rho_{dm})},
\end{equation}
where $b$ is a constant, $H$ the Hubble parameter, and $\rho_{de}$ and $\rho_{dm}$ are the energy densities of dark energy and dark matter, respectively. On the other hand, interactions such as $Q = 3 H b \rho_{de}$, $Q = 3 H b \rho_{dm}$ and $Q = 3 H b (\rho_{de} + \rho_{de})$ can completely suppress future finite-time singularity formation, for high values of $b$. On the other hand, for some regions of the parameter space the mentioned interactions do not affect the type of singularity. A similar conclusion has been achieved if we consider $Q = 3 b H \rho_{de}\rho_{dm}/(\rho_{de}+\rho_{dm})$ and $Q = 3 b H \rho_{de}^{2}/(\rho_{de}+\rho_{dm})$ interactions. The non-gravitational interaction in this paper is understood as in other similar papers, i.e., it is assumed that the non-gravitational interaction is responsible for dark energy transfer to dark matter, in the following way
\begin{equation}\label{eq:INTDE}
\dot{\rho}_{de} + 3 H (\rho_{de} + P_{de}) = -Q,
\end{equation}
and
\begin{equation}\label{eq:INTDM}
\dot{\rho}_{dm} + 3 H \rho_{dm} = Q,
\end{equation}
where $Q$ stands for the non-gravitational interaction, and dark matter is assumed to be pressureless. The background dynamics, according to Friedmann's equation, is
\begin{equation}
\frac{\ddot{a}}{a} = -\frac{4 \pi G}{3} (\rho_{eff} + P_{eff}),
\end{equation}
where $a$ is the scale factor, while $\rho_{eff} = \rho_{de} + \rho_{dm}$ and $P_{eff} = P_{de}$. The last equation, together with Eqs.~(\ref{eq:INTDE}), (\ref{eq:INTDM}) and (\ref{eq:Waals}), for appropriate initial conditions, allow to perform the study of the model. We would like to mention that the singularities occurring in Refs.~\cite{S4} and \cite{S5} have not been observed during the presented study.

In what follows, the $Om$ analysis has been used in order to estimate the phantom nature of the model, taking into account that the analysis is a null test for the $\Lambda$CDM model with $Om_{\Lambda CDM} = \Omega^{(0)}_{dm}$.
The two-point diagnostics $Om(z_{i}, z_{j})$ and $Omh^{2}(z_{i}, z_{j})$~(extensions of the $Om$ analysis) have been introduced for the same purposes as the $Om$ analysis itself. These diagnostics have been used intensively to study various cosmological models. On the other hand, the recent results of Ref.~\cite{Om} with $N = 36$ measurements of H(z) from BAOs and the differential ages~(DAs) of passively evolving galaxies, support the claim that $\Lambda$CDM is in tension with H(z) data, according to the two-point diagnostics developed by Shafieloo, Sahni, and Starobinsky. Moreover, as stated in Ref.~\cite{Om}, serious systematic differences between the BAO and DA methods should be understood before $H(z)$ measurements can compete with other probe methods. Keeping the mentioned results in mind, we should use $Om$ and $Omh^{2}(z_{i}, z_{j})$ diagnostics for the models in question, accepting that different trajectories, for instance, on $Om -z$ plane could actually signal the differences between the models.

This paper is organized as follows. In Sect.~\ref{sec:CC} the family of models with linear non-gravitational interacting van der Waals dark fluids is introduced. We use the graphical behavior of the parameters to describe how the non-gravitational interactions change the background dynamics. $Om$ analysis and constraints known from the $Omh^{2}$ analysis are also employed to determine finite-time future singularities. Moreover, in Sect.~\ref{sec:CCC} a similar study is carried out for three different examples of non-linear non-gravitational interactions. In total, six different forms of non-gravitational interactions are thus considered in the paper. Only one of them is seen to be able to change the type of the singularity arising in the non-interacting model. It is quite interesting to observe that, for appropriate values of the parameters, all of the examples of interactions here considered can suppress finite-time future singularity formation. Finally, the last section, Sect.~\ref{sec:D}, is devoted to summarize all the results obtained, to conclusions and an outlook.

\section{Singularities with linear non-gravitational interactions}\label{sec:CC}

\begin{figure}[h!]
\begin{center}$
\begin{array}{cccc}
\includegraphics[width=100 mm]{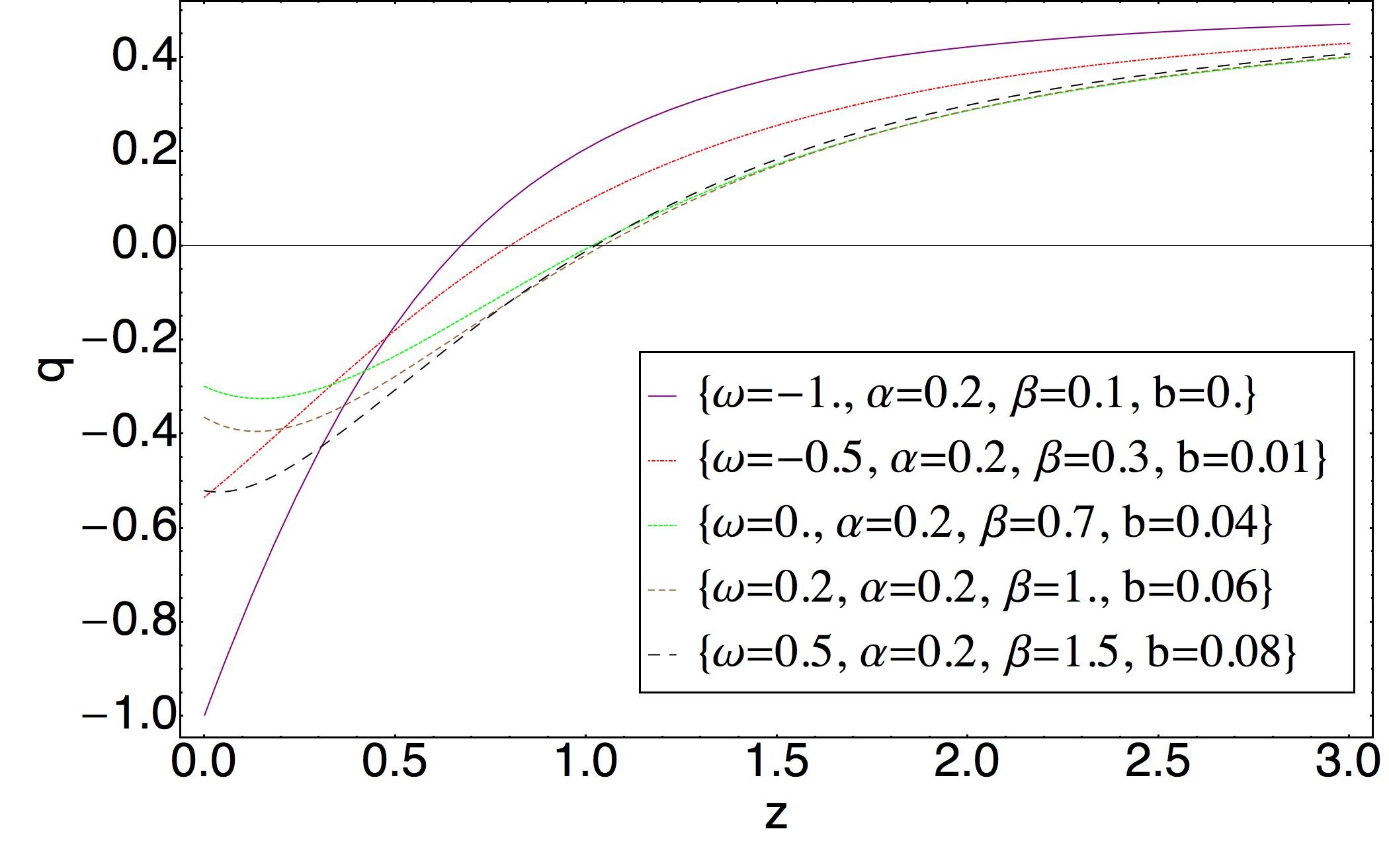} &
\end{array}$
\end{center}
\caption{Plot of the behavior of the deceleration parameter $q$ for the interacting three-parameter van der Waals dark fluid universe, when the non-gravitational interaction is given by Eq.~(\ref{eq:QM1}). We set $H_{0} = 0.69$ and $\Omega^{(0)}_{dm} = 0.27$.}
\label{fig:1}
\end{figure}

\begin{figure}[h!]
\begin{center}$
\begin{array}{cccc}
\includegraphics[width=85 mm]{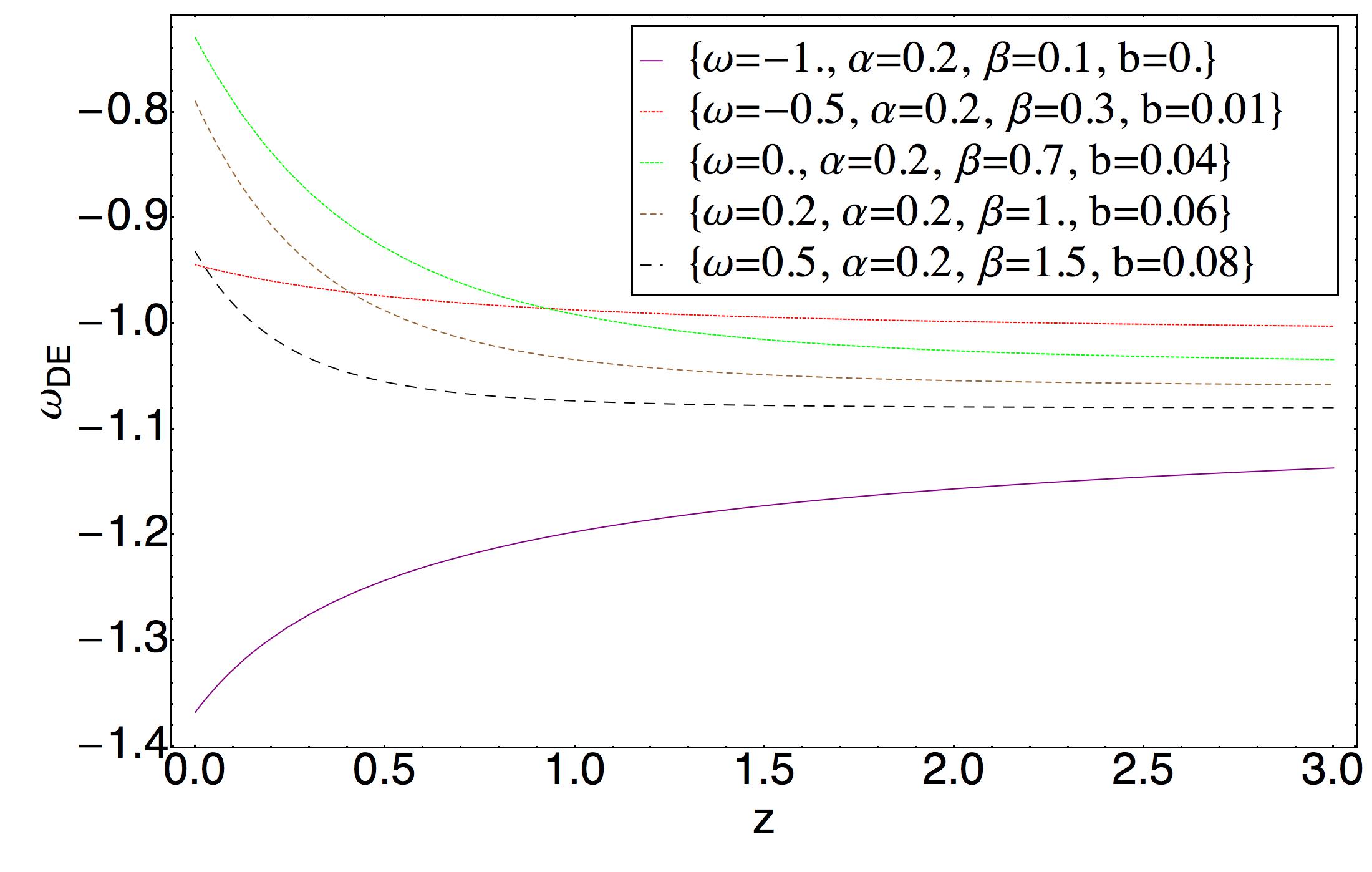} &
\includegraphics[width=85 mm]{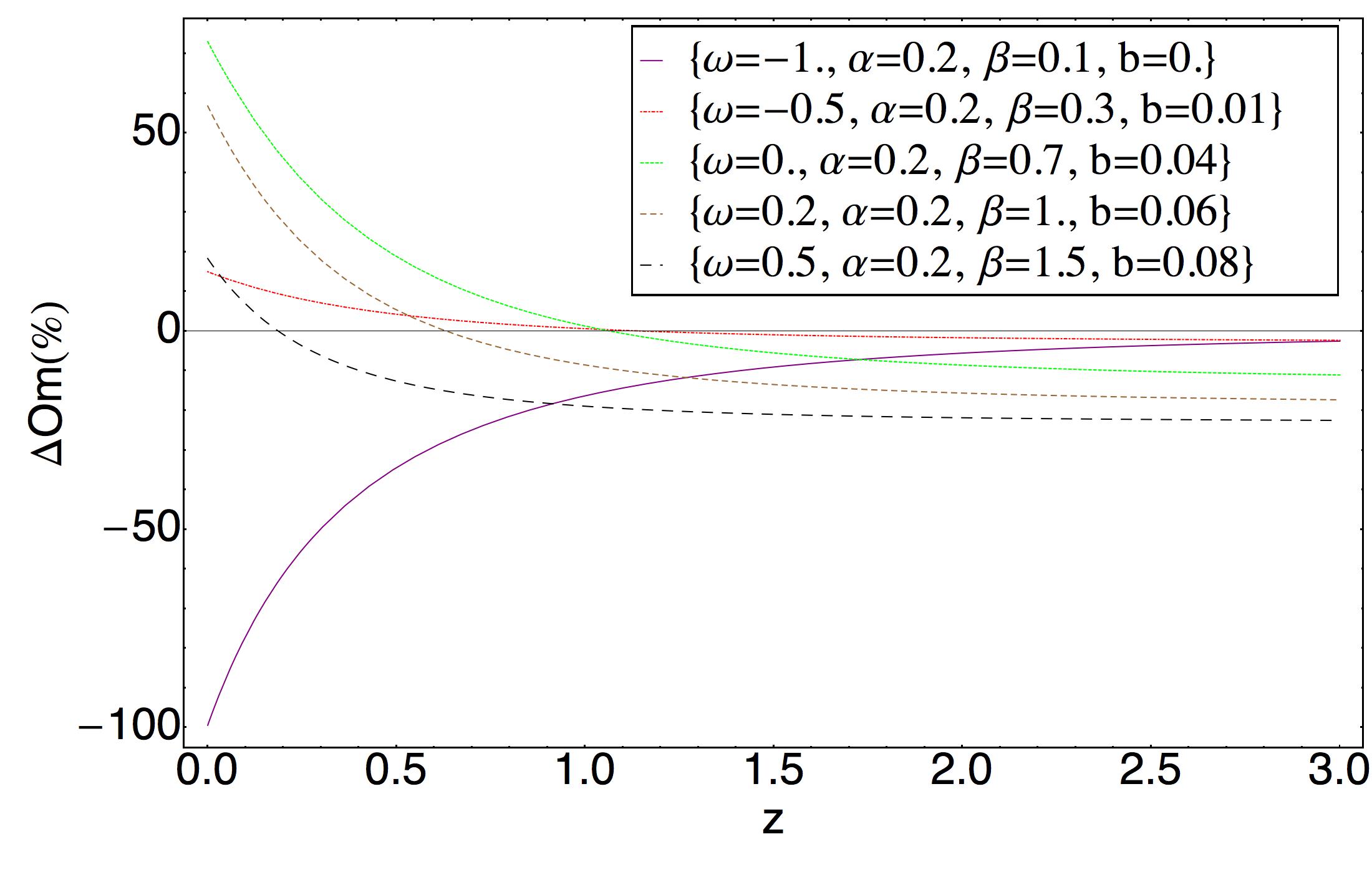}
\end{array}$
\end{center}
\caption{Plot of the behavior of the EoS parameter $\omega_{de}$ corresponding to the dark fluid described by Eq.~(\ref{eq:Waals}) (left). Behavior of $\Delta Om$ of the model (right). The non-gravitational interaction is given by Eq.~(\ref{eq:QM1}). We set $H_{0} = 0.69$ and $\Omega^{(0)}_{dm} = 0.27$.}
\label{fig:2}
\end{figure}

In this section we will explore some crucial aspects of the low redshift universe resulting for different  forms of the singularity-fixed linear non-gravitational interaction between the van der Waals dark fluid and dark matter. An example of interaction of this kind can be constructed using, for instance, the Hubble parameter $H$ and the dark energy  density $\rho_{de}$, in the following way
\begin{equation}\label{eq:QM1}
Q = 3 H b \rho_{de},
\end{equation}
where $b$ is a constant. Such interaction, for $b > 0$, guarantees a dark energy flow to dark matter during the evolution of the low redshift universe. A study of the behavior of the deceleration parameter $q$, when the non-gravitational interaction is given by Eq.~(\ref{eq:QM1}), shows that a growth of $b$ will significantly increase the transition redshift $z_{tr}$, as compared to the non-interacting model~($z_{tr} \approx 0.6$) represented by the solid purple curve in Fig.~(\ref{fig:1}). Moreover, we have seen that the value of the deceleration parameter $q$ at $z=0$ is not affected. On the other hand, on Fig.~(\ref{fig:1}) we see the plot representing the graphical behavior of the same parameter for different values of $\omega$, $\beta$, and $b$, for fixed $\alpha$ and for $H_{0} = 0.69$ and $\Omega^{(0)}_{dm} = 0.27$. We observe  that the parameters of the model  have strong impact on the behavior of the deceleration parameter $q$ and on that of the transition redshift $z_{tr}$, too. In particular, two distinct behaviors of the deceleration parameter could be observed for low redshifts, mainly due to the change of the $\beta$ parameter. Namely, the deceleration parameter either will be only decreasing, thus preserving the accelerated expanding phase in the future, or either, starting from some redshift, it will grow and in the future an exit from the accelerated phase will be observed. On the other hand, the two plots of Fig.~(\ref{fig:2}) allow us to understand some aspects of the evolution of the van der Waals dark fluid, in terms of EoS parameter, and to estimate the deviation of the model from standard $\Lambda$CDM, in terms of the parameter $\Delta Om$,
\begin{equation}
\Delta Om (\%) = 100 \times \left [ \frac{ Om_{Model} }{ Om_{\Lambda CDM}} -1 \right ],
\end{equation}
in universes where the acceleration of the background is given in Fig.~(\ref{fig:1}). First, it is seen that, in the case of a non-interacting model, the van der Waals fluid has only a phantom nature. Moreover, the study shows~(as seen  on the left plot of Fig.~(\ref{fig:2})) that, at higher redshifts, the  three-parameter van der Waals dark fluid considered could have a phantom~($\omega_{de} < -1$) or quintessence~($\omega_{de} > -1$) nature, but that at lower redshifts it will develop a quintessence nature, only.

Moreover, the observed growth of the deceleration parameter $q$ for some values of the parameters of the model  is due to an increase of $\omega_{DE} = P_{de}/\rho_{de}$. In general, our study shows that the $Om$ analysis is a good tool in order to quantify the fitness of the model and clearly allows to estimate the deviations from the $\Lambda$CDM model with $Om=0.27$. The values of the parameters here considered arise from the constraints coming from the $Omh^{2}$ analysis for $z=0$, $z=0.57$, and $z=2.34$, respectively. One of the most interesting aspects observed in this model is the clear possibility to reproduce the $\Lambda$CDM  behavior for higher redshifts while evolving to a quintessence solution for lower redshifts. As a general trend, a detailed study shows that the ($\omega$, $\alpha$, $\beta$, $b$) parameter space, for fixed $H_{0}$ and $\Omega^{0}_{dm}$, can be split into various regions, allowing to produce examples of background evolution which are well in agreement with the most recent observational data available, and to predict moreover interesting possible future evolutions of the toy model. This directly affects the type of  future singularity one is going to encounter, and also the singularity forming time.

The characteristic future singularity for the non-interacting model is of type IV, which has been observed, for instance, with $\omega = -1$, $\alpha = 0.2$, $\beta = 0.1$ and $b=0.0$. Later, the interaction Eq.~(\ref{eq:QM1}) with an increasing value  of the parameter $b$ will just delay the singularity formation time without ever changing the type of the singularity itself. On the other hand, Table~\ref{tab:Table1} shows values of the parameters for which the corresponding interacting model has a singularity-free future evolution.

\begin{table}
  \centering
    \begin{tabular}{ | l | l | l | l | p{4cm} |}
    \hline
 $\omega$ & $\alpha$ & $\beta$ & $b$ & Expansion in future \\
      \hline
  $-0.5$ & $0.2$ & $0.3$ & $0.01$ & accelerated\\
          \hline
 $0.0$ & $0.2$ & $0.7$ & $0.04$ & accelerated\\
    \hline
 $0.2$ & $0.2$ & $1.0$ & $0.06$ & accelerated\\
     \hline
  $0.5$ & $0.2$ & $1.5$ & $0.08$ & decelerated\\
     \hline

    \end{tabular}
\caption{Values of the model parameters providing an interacting three-parameter van der Waals dark fluid universe with a singularity-free future evolution. The non-gravitational interaction is given by Eq.~(\ref{eq:QM1}). Here  $H_{0} = 0.69$ and $\Omega^{(0)}_{dm} = 0.27$.}
  \label{tab:Table1}
\end{table}

\begin{figure}[h!]
\begin{center}$
\begin{array}{cccc}
\includegraphics[width=85 mm]{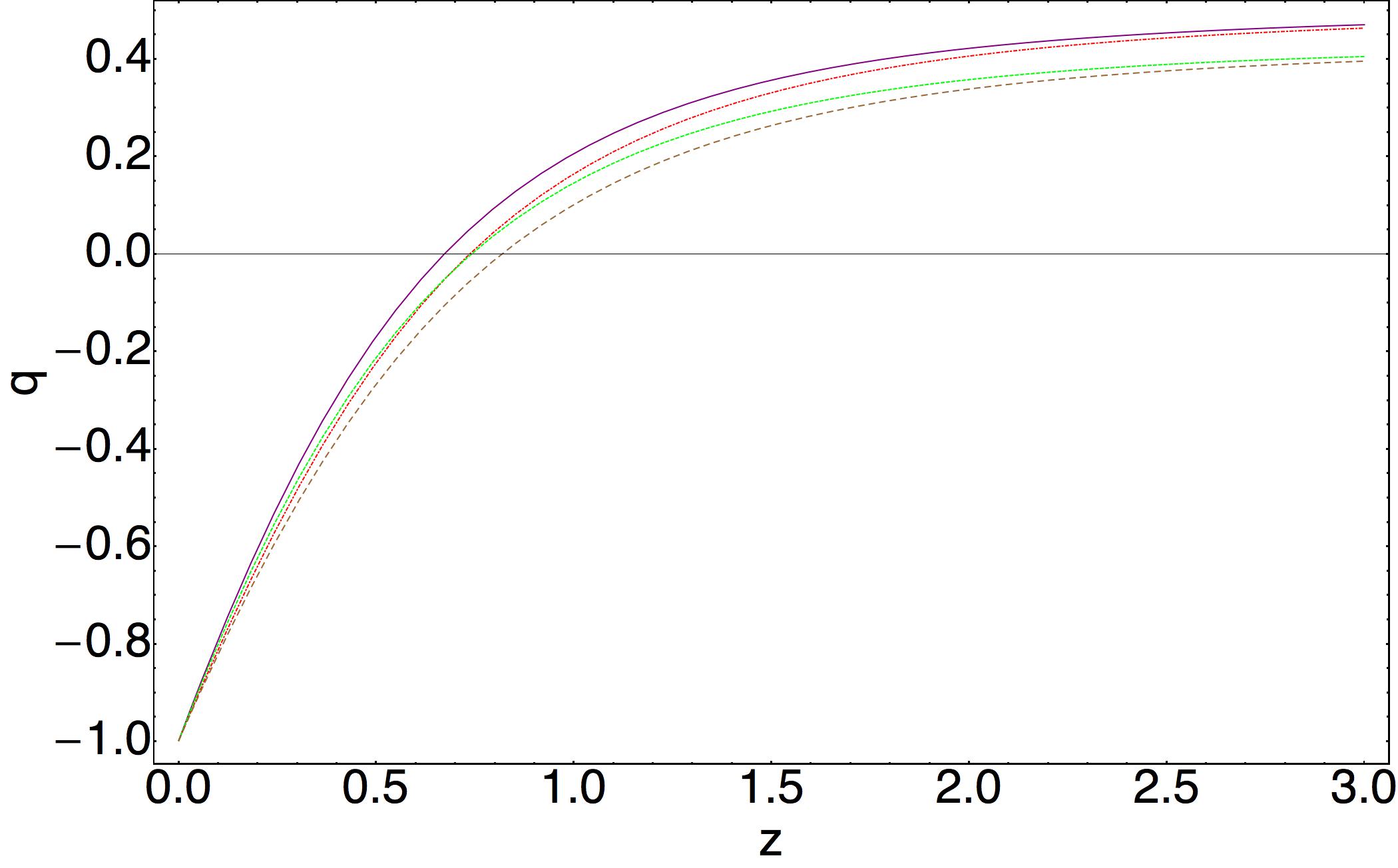}&
\includegraphics[width=85 mm]{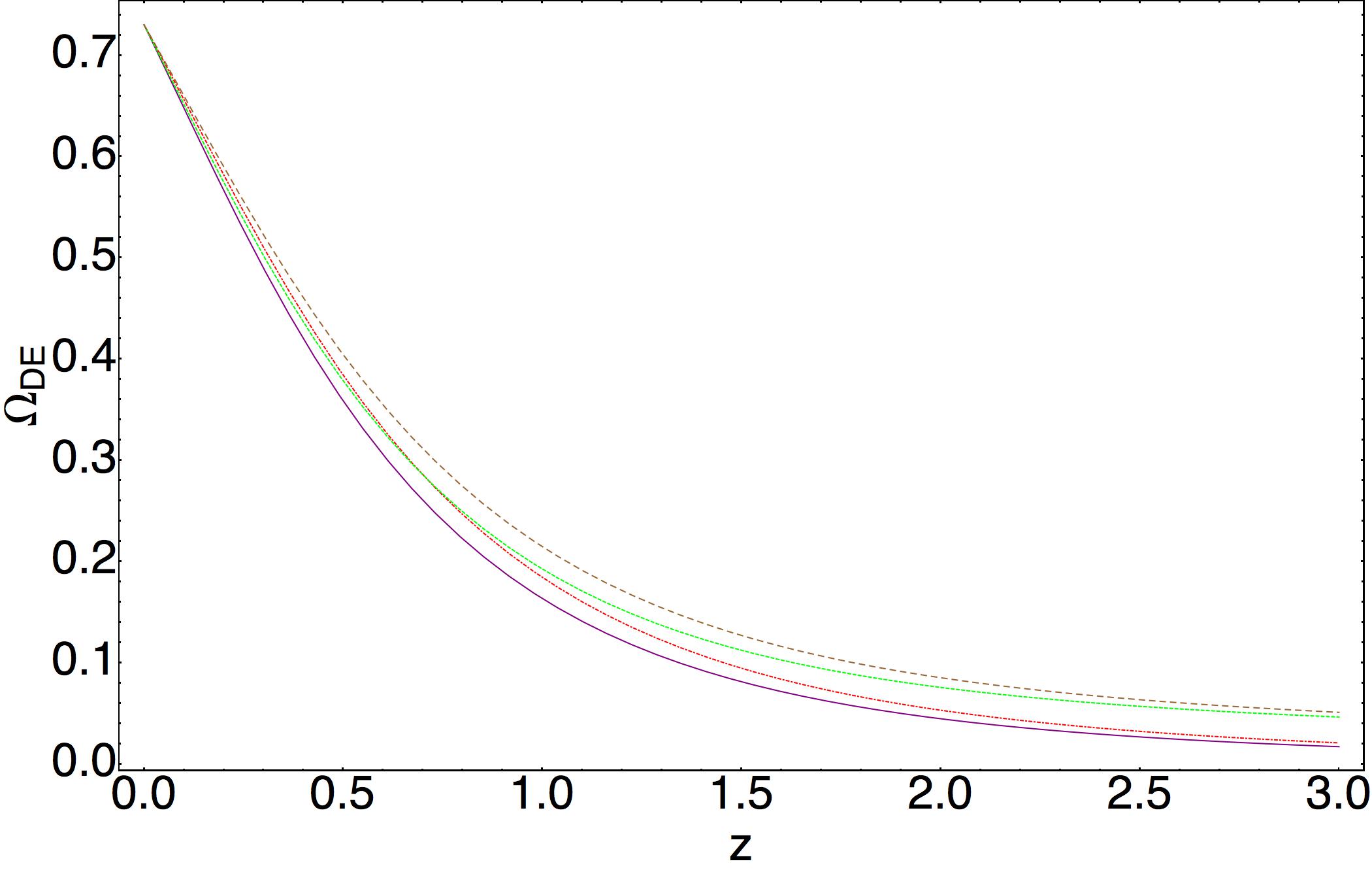}
\end{array}$
\end{center}
\caption{Behavior of the deceleration parameter $q$ and of $\Omega_{de}$ for an interacting three-parameter van der Waals dark fluid universe. The purple curve corresponds to a non-interacting model, while the red, green and brown curves represent interacting models with non-gravitational interactions given by Eq.~(\ref{eq:QM1}), Eq.~(\ref{eq:QM2}) and Eq.~(\ref{eq:QM3}), respectively. Here $\omega = -1.0$, $\alpha = 0.2$, $\beta = 0.1$, $H_{0} = 0.69$, $\Omega^{(0)}_{dm} = 0.27$, and $b = 0.04$.}
\label{fig:3}
\end{figure}

\begin{figure}[h!]
\begin{center}$
\begin{array}{cccc}
\includegraphics[width=85 mm]{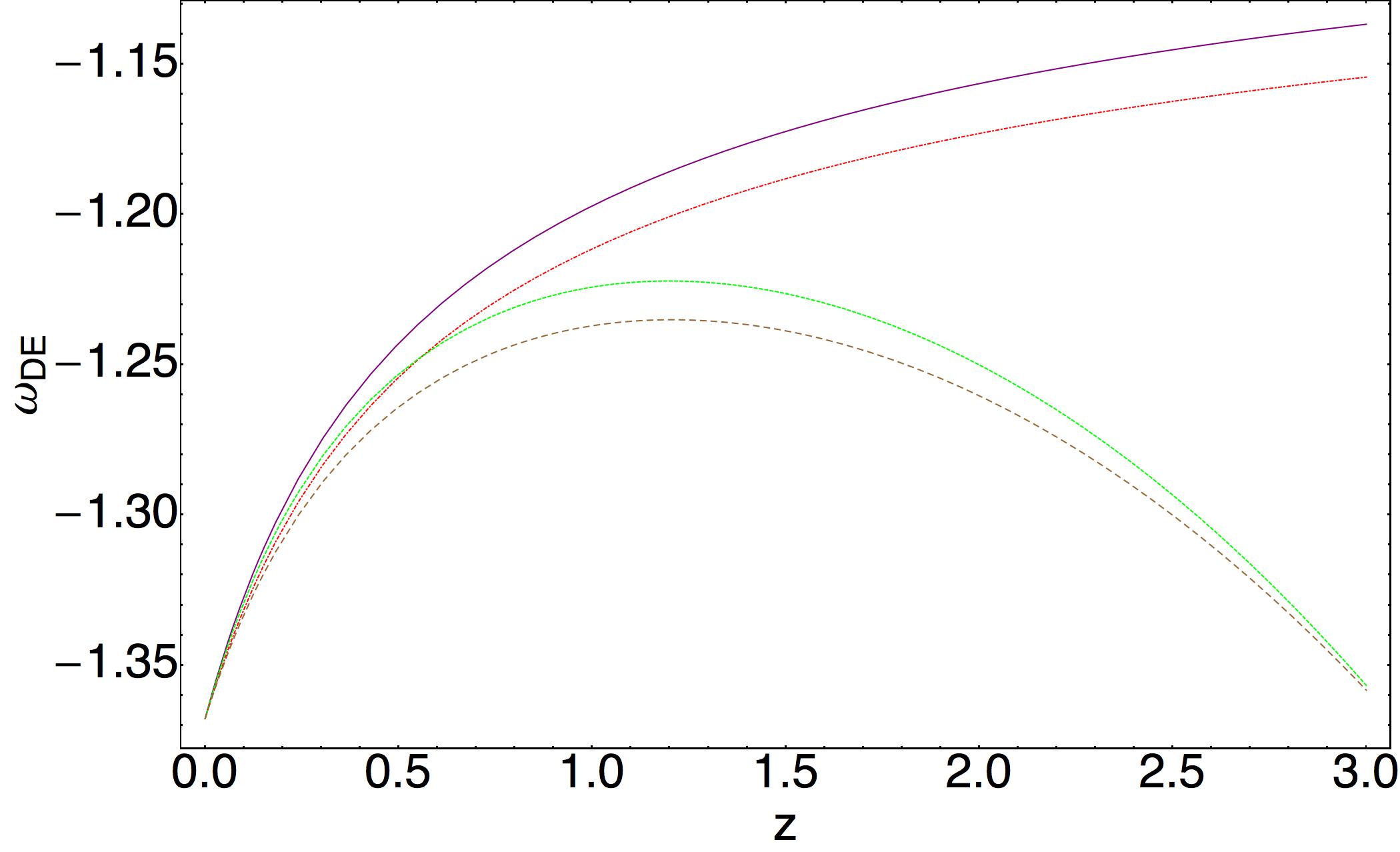}&
\includegraphics[width=85 mm]{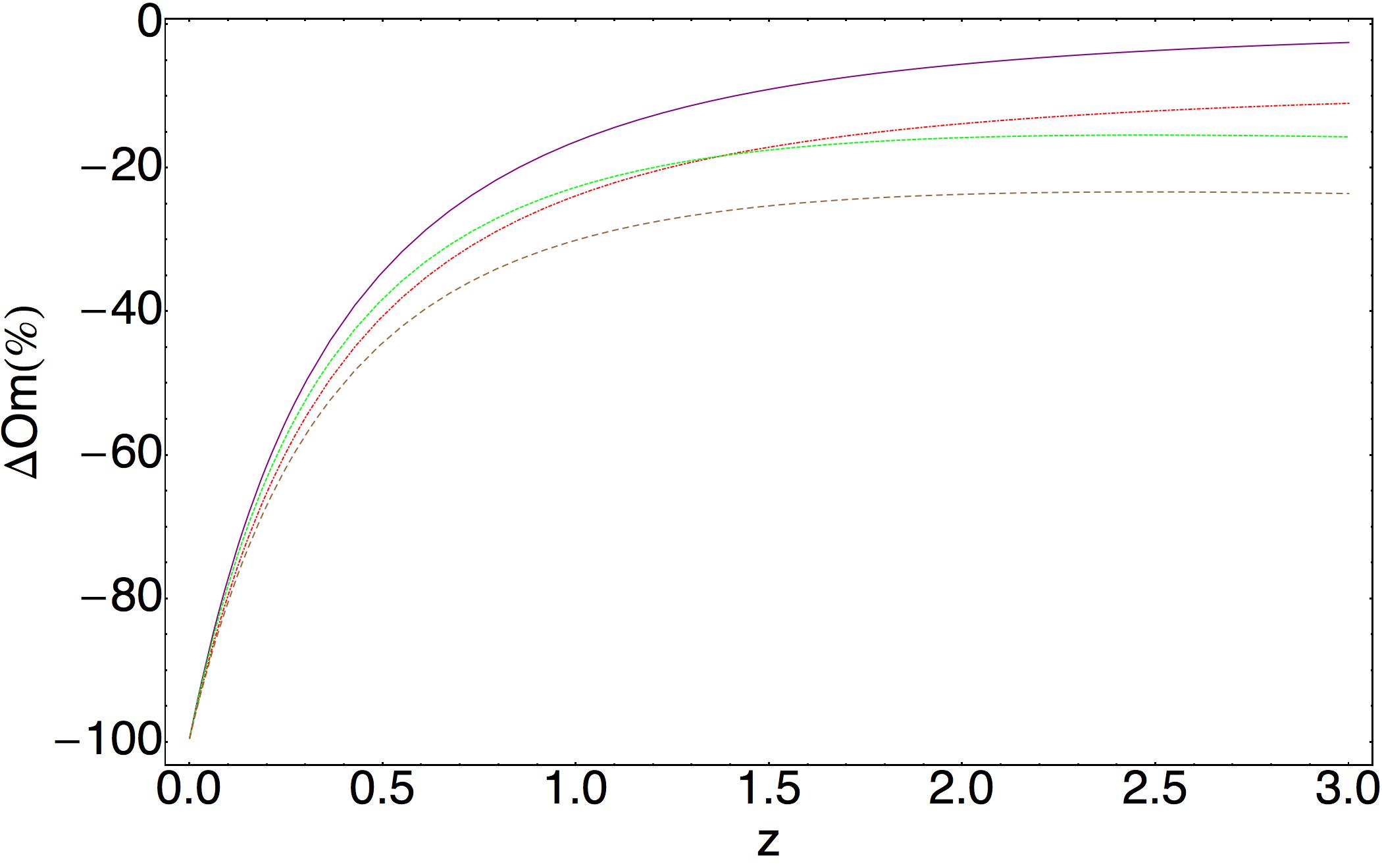}
\end{array}$
\end{center}
\caption{Behavior of the deceleration parameter $q$ for an interacting three-parameter van der Waals dark fluid universe. The purple curve corresponds to a non-interacting model, while the red, green and brown curves represent interacting models with non-gravitational interactions given by Eq.~(\ref{eq:QM1}), Eq.~(\ref{eq:QM2}), and Eq.~(\ref{eq:QM3}), respectively. Here $\omega = -1.0$, $\alpha = 0.2$, $\beta = 0.1$, $H_{0} = 0.69$, $\Omega^{(0)}_{dm} = 0.27$, and $b = 0.04$.}
\label{fig:3_1}
\end{figure}

Finally, to complete a comprehensive picture of the model, we have assumed the following two forms for the non-gravitational interaction:
\begin{equation}\label{eq:QM2}
Q = 3 H b \rho_{dm},
\end{equation}
and
\begin{equation}\label{eq:QM3}
Q = 3 H b (\rho_{de} + \rho_{dm}),
\end{equation}
which do not change the type of the singularity already observed in the case of the non-interacting models. They will just affect the singularity formation time. At the end of this section, in Fig.~(\ref{fig:3}) we compare the impact of the  non-gravitational interactions here considered on the deceleration parameter $q$ and on $\Omega_{de}$, while Fig.~(\ref{fig:3_1}) exhibits the impact on the EoS parameter of the three-parameter van der Waals dark fluid and $\Delta Om$. The left plot in Fig.~(\ref{fig:3}) clearly shows that the non-gravitation interaction  $Q = 3 H b (\rho_{de} + \rho_{dm})$, as compared with other forms of non-gravitational interactions, gives a higher values for the transition redshift, i.e. in this case the transition will take place earlier, as compared with other cases. On the other hand, the transition redshift is the same for the interactions $Q = 3 H b \rho_{de}$ and $Q = 3 H b \rho_{dm}$. Finally, the left plot of Fig.~(\ref{fig:3_1}) proves that $Q = 3 H b (\rho_{de} + \rho_{dm})$ and $Q = 3 H b \rho_{dm}$ provide the smallest value for $\omega_{de}$. Moreover, the phantom nature of the three-parameter van der Waals fluid for higher and lower redshifts is apparent. The right plot in Fig.~(\ref{fig:3_1}) shows that, for higher redshifts $Q = 3 H b (\rho_{de} + \rho_{dm})$, the interaction provides a stronger deviation from the $\Lambda$CDM model, which will further grow with decreasing redshift. Eventually, for lower redshifts~($z<0.15 )$ the observed deviation is not affected for the  linear non-gravitational interactions here considered.

\section{Singularities with non-linear, non-gravitational interactions}\label{sec:CCC}
In this section we address  the issue of the  classification of future finite-time singularities in the case of the three non-linear, non-gravitational interactions given in Eq.~(\ref{eq:QM23}),
\begin{equation}\label{eq:QM21}
Q = 3 b H \frac{\rho_{de}^{2}}{(\rho_{de}+\rho_{dm})},
\end{equation}
and
\begin{equation}\label{eq:QM22}
Q = 3 b H \frac{\rho_{de}\rho_{dm}}{(\rho_{de}+\rho_{dm})}.
\end{equation}
First of all, we observe that the non-gravitational interactions, Eqs.~(\ref{eq:QM21}) and (\ref{eq:QM22}), will not change the type of the singularity already formed in the case of a non-interacting model. On the other hand, they will strongly affect  the precise time the singularity forms. In particular, with an increase of $b$ singularity formation will be delayed. In other words, the qualitative picture is similar to the cases considered in the previous section with linear non-gravitational interactions. 

However, the situation will change drastically if we consider a non-gravitational interaction, as  the one  given by Eq.~(\ref{eq:QM23}), and we see from the corresponding picture that this is the most interesting form of non-gravitational interaction among those considered in this paper, because it can change the type of the singularity which was formed in the case of the non-gravitational model. Specifically, we observe that it changes the   Type IV singularity to a Type II one. On the other hand, the parameter space contains regions, which provide singularity-free future evolution for the models with the non-linear, non-gravitational interactions here considered, what is quite remarkable. In these cases we will just have an expanding universe accelerating for ever. 

For the two types of models considered in this paper, we have also studied the behavior of $C^{2}_{s}$ and have seen that $b>1.0$  will produce, in all cases, an effective fluid which is not stable at low redshifts. In other words, we have observed that the formation of the singularity occurs for $C^{2}_{s} > 1$, while a singularity-free evolution takes place when $C^{2}_{s} < 0$ at low redshifts~($z< 0.3$). However, it is also possible to have a stable effective fluid, for a very tiny value of the interaction parameter $b$, and get a singularity-free evolution of the future universe.

\section{Discussion and conclusions}\label{sec:D}

In this paper we have aimed at filling an important gap that existed in studies of cosmological models with a van der Waals dark fluid playing the role of dark energy; more specifically, when this dark fluid is interacting, as compared with the non-interacting case, and paying special attention to the differences between the possible future singularities in each situation. In particular, we have explicitly considered  six different forms of non-gravitational interaction between the three-parameter van der Waals dark fluid and cold dark matter; the corresponding future finite-time singularities have been found and classified. Moreover, the evolution of these singularities for varying values of the parameters of the models have been discussed in detail.

Our study shows that the parameter space can be conveniently divided into various regions and that, for some of those regions, the non-gravitational interactions of the type $Q = 3 H b \rho_{de}$, $Q = 3 H b \rho_{dm}$ and $Q = 3 H b (\rho_{de} + \rho_{de})$ can completely suppress future finite-time singularity formation, for high values of $b$. On the other hand, for some of these regions in the parameter space, the above mentioned interactions do not affect the type of singularity, which was already present in the non-interacting case. More specifically, the Type IV singularity already occurring in the case of the non-interacting model is preserved by the interaction.

A similar conclusion has been archived when considering the following cases of non-gravitational interaction: $Q = 3 b H \rho_{de}\rho_{dm}/(\rho_{de}+\rho_{dm})$, $Q = 3 b H \rho_{dm}^{2}/(\rho_{de}+\rho_{dm})$, and $Q = 3 b H \rho_{de}^{2}/(\rho_{de}+\rho_{dm})$, with the only one difference arising in the case of the interaction $Q = 3 b H \rho_{dm}^{2}/(\rho_{de}+\rho_{dm})$, which is namely able to change the Type IV singularity of the corresponding non-interacting model to a Type II~(The Sudden) singularity. Moreover, during the study of $C^{2}_{s}$ we have found that the violation of the constraints for the effective fluid in this case could actually be one of the reasons for the formation of the finite-time singularity.

Some aspects of the respective impacts of the three models above, for non-gravitational interactions, on the deceleration parameter $q$, and on $\omega_{de}$, $\Omega_{de}$ and $Om$, have been also discussed in much detail. We have proved, in particular, that the $Om$ analysis can be successfully carried out for a proper study of these models. On the other hand, we have found that if we take altogether the thirty-six measurements presented in the Introduction, then the $Omh^{2}$ analysis  does not practically affect the constraints obtained from the same analysis with only three measurements, as subsequently used in this paper. Of course, for a full analysis we need to involve several data sets including, in particular, strong and weak gravitational lensing data, in order to get more precise constraints on the parameters of the model. However, we are quite certain that the result presented here will not be much affected in this case, owing to the fact that we have scanned quite extensive domains of the whole parameter region in the course of our study.

In summary, the present analysis has recovered various interesting aspects of the van der Waals dark energy universe revealing at the same time some new and very promising directions. With the use of our simple model we have demonstrated that a non-gravitational interaction can sometimes completely suppress the formation of future finite-time singularities, or either it can change the type of the singularity. Moreover, the non-gravitational interaction of the van der Waals dark fluid can strongly affect the singularity formation time. It is also possible to suppress the singularity that forms in the non-interacting model. 

It would be interesting to extend the present study in order to involve new aspects of the interacting dark energy models, which have recently appeared in the literature. In particular, it would be important to see how sign-changing interactions can influence the type of the singularity. And it would also be interesting to extend the present study to take into account viscosity, in the light of a recent and quite revealing discussion in~\cite{W3} in relation to the van der Waals inflationary universe. All these aspects will be considered in subsequent work.

\medskip

\noindent {\bf Acknowledgements}.  EE was partially supported by CSIC,
I-LINK1019 Project, by MINECO (Spain), Projects FIS2013-44881-P and
FIS2016-76363-P, and by the CPAN Consolider Ingenio Project, and was also
partially supported by JSPS (Japan), Fellowship N. S17017 (E. Elizalde). The
paper was started while EE was visiting the Kobayashi-Maskawa Institute
in Nagoya, Japan. He is much obliged with S. Nojiri and the rest of the members
of the KMI for the very warm hospitality.


\begin{thebibliography}{9}

\bibitem{Od1}
S. Nojiri, S.D. Odintsov, V.K. Oikonomou, Phys. Rept. 692, 1-104 (2017), DOI:10.1016/j.physrep.2017.06.001 [arXiv:1705.11098 [gr-qc]]; Jaewon Yoo, Yuki Watanabe, Int. J. Mod. Phys. D 21, 1230002 (2012), DOI:10.1142/S0218271812300029 [arXiv:1212.4726 [astro-ph.CO]]; T. Clifton et al., Physics Reports 513, 1, 1-189 (2012), DOI:10.1016/j.physrep.2012.01.001 [arXiv:1106.2476 [astro-ph.CO]]; S. Nojiri and S. D. Odintsov, Int. J. Geom. Methods Mod. Phys. 04, 115 (2007); S. Nojiri and S. D. Odintsov, Phys. Rept. 505:59-144 (2011); S. Capozziello et al, Phys. Lett. B 639:135-143 (2006); Yi-Fu Cai et al, Rept. Prog. Phys. 79, no. 4, 106901 (2016); S. Capozziello, M. De Laurentis, Physics Reports 509, 167321 (2011); E.~Elizalde, S.~Nojiri, S.~D.~Odintsov and D.~Saez-Gomez, Eur.\ Phys.\ J.\ C {\bf 70}, 351S (2010); E.~Elizalde, R.~Myrzakulov, V.~V.~Obukhov and D.~Saez-Gomez, Class.\ Quant.\ Grav.\  {\bf 27}, 095007 (2010); Nojiri and S. D. Odintsov, Phys. Rev. D 68, 123512 (2003); G.~Cognola, E.~Elizalde, S.~Nojiri, S.~D.~Odintsov, L.~Sebastiani and S.~Zerbini, Phys.\ Rev.\ D {\bf 77}, 046009 (2008); F.~Briscese, E.~Elizalde, S.~Nojiri and S.~D.~Odintsov, Phys.\ Lett.\ B {\bf 646}, 105 (2007); E.~Elizalde, S.~Nojiri and S.~D.~Odintsov, Phys.\ Rev.\ D {\bf 70}, 043539 (2004); I.~Brevik, E.~Elizalde, S.~Nojiri and S.~D.~Odintsov, Phys.\ Rev.\ D {\bf 84}, 103508 (2011); S. Nojiri and S. D. Odintsov, Phys. Rev. D 74, 086005 (2006); V.K. Oikonomou, E. N. Saridakis, Phys. Rev. D 94, 124005 (2016)

\bibitem{Od3}
Francisco S. N. Lobo, Class.Quant.Grav. 23, 1525-1541 (2006), DOI:10.1088/0264-9381/23/5/006 [arXiv:0508115[gr-qc]]; P. Brax et al., Phys. Rev. D 95, 083514 (2017), DOI:10.1103/PhysRevD.95.083514 [arXiv:1702.02983 [gr-qc]]; S. H. Hendi et al., JCAP 09, 013 (2016), DOI: 10.1088/1475-7516/2016/09/013 [arXiv:1509.05145 [hep-th]]; A. V. Astashenok et al., JCAP 12, 040 (2013), DOI:10.1088/1475-7516/2013/12/040 [arXiv:1309.1978 [gr-qc]]; A. V. Astashenok et al, Class. Quantum Grav. 34, 205008 (2017), DOI:10.1088/1361-6382/aa8971 [arXiv:1704.08311 [gr-qc]]; A. V. Astashenok et al, JCAP 01, 001 (2015), DOI:10.1088/1475-7516/2015/01/001 [arXiv:1408.3856 [gr-qc]]; M. Zaeem-ul-Haq Bhatti et al, Mod. Phys. Lett. A 32, 1750042 (2017); Amit Das et al, Eur. Phys. J. C  76:654 (2016), DOI:	10.1140/epjc/s10052-016-4503-0 [	arXiv:1608.00566 [gr-qc]]; D. Momeni, et al,  arXiv:1611.03727 [gr-qc]; Hector O. Silva et al,  Int. J. Mod. Phys. D 25, 1641006 (2016) 	 DOI:10.1142/S0218271816410066 [arXiv:1602.05997 [gr-qc]]; S.~Capozziello, V.~F.~Cardone, E.~Elizalde, S.~Nojiri and S.~D.~Odintsov, Phys.\ Rev.\ D {\bf 73}, 043512 (2006).

\bibitem{M1}
J. Sadeghi et al, JCAP 12, 031 (2013); J. Sadeghi et al, Int J Theor Phys 53: 2246 (2014); J. Sadeghi et al, RAA Vol. 15 No. 2, 175190 (2015); J. Sadeghi et al, Int. J. Mod. Phys. D 25, 1650108 (2016); J. Sadeghi et al, Advances in High Energy Physics, Article ID 129085 (2014); M. Khurshudyan et al, Int J Theor Phys 53: 2370 (2014); M. Khurshudyan et al, Astrophys Space Sci 357: 113 (2015); M. Khurshudyan, R Myrzakulov, The European Physical Journal C 77 (2), 65 (2017) ; M. Khurshudyan, Astrophys Space Sci 360: 33 (2015); M. Khurshudyan, Astrophys Space Sci 361: 232 (2016); I. Brevik et al, arXiv:1706.02543; E.~Elizalde, S.~D.~Odintsov, L.~Sebastiani and R.~Myrzakulov, Nucl.\ Phys.\ B {\bf 921}, 411 (2017); M. Zh. Khurshudyan and A. N. Makarenko, Astrophys Space Sci 361: 187 (2016); M Khurshudyan, Eur. Phys. J. Plus 131: 25 (2016).

\bibitem{M2}
M. Khurshudyan, Mod. Phys. Lett. A, 31, 1650055 (2016); M. Khurshudyan, Mod. Phys. Lett. A, 31, 1650097 (2016); M. Khurshudyan, Symmetry, 8(11), 110 (2016); S. Nojiri, S. Odintsov, Gen Relativ Gravit 38: 1285 (2006) ; S. Nojiri and S. D. Odintsov, arXiv: 1703.06372; E.~Elizalde and L.~G.~T.~Silva, Astrophys.\ Space Sci.\  {\bf 362}, no. 1, 7 (2017); I.~Brevik, E.~Elizalde, V.~V.~Obukhov and A.~V.~Timoshkin, Annalen Phys.\  {\bf 529}, 0195 (2017);  E. N. Saridakis, Phys. Lett. B 676:7-11 (2009); Kazuharu Bamba, Sergei D. Odintsov, Symmetry 2015, 7, 220-240; S. Nojiri et al, arXiv:1705.11098; S. Chattopadhyay et al, Eur. Phys. J. C 74:3080 (2014)


\bibitem{W1}
G. M. Kremer, Cosmological models described by a mixture of van der Waals fluid and dark energy, Phys. Rev. D 68, 123507 (2003) [arXiv:0309111[gr-qc]]


\bibitem{W2}
G. Vardiashvili, E. Halstead, R. Poltis, A. Morgan, D. Tobar, Inflationary Constraints on the Van der Waals Equation of State, arXiv:1701.00748 [gr-qc]


\bibitem{W3}
I.~Brevik, E.~Elizalde, S.~D.~Odintsov and A.~V.~Timoshkin, Int. J. Geom. Meth. Mod. Phys. {\bf 14}, 1750185 (2017), DOI: 10.1142/S0219887817501857, arXiv:1708.06244 [gr-qc].


\bibitem{W4}
G.M. Kremer, Gen. Relativ. Gravit. 36, 1423 (2004); S. Capozziello et al,  JCAP 0504:005 (2005), DOI:10.1088/1475-7516/2005/04/005 [arXiv:astro-ph/0410503]; Rudinei C. S. Jantsch et al, Int. J. Mod. Phys. D 25, 1650031 (2016), https://doi.org/10.1142/S0218271816500310	[arXiv:1601.05337 [gr-qc]]; S. Capozziello et al, Phys. Lett. A 299 (2002) 494; G. M. Kremer, Phys. Rev. D 68, 123507 (2003); S. Capozziello et al, Recent Res. Dev. Astron. Astrophys. 1, 625 (2003);
S. Capozziello et al, JCAP 04, 005 (2005); V. F. Cardone et al, Phys. Rev. D 73, 043508 (2006); B. Saha, Int. J. Theor. Phys. 45, 952 (2006); M. Khurshudyan, B. Pourhassan and E. O. Kahya, Int. J. Geom. Methods Mod. Phys. 11 1450061 (2014)

\bibitem{S1}
S. Nojiri, S. D. Odintsov, S. Tsujikawa, Phys. Rev. D 71:063004 (2005) [arXiv:hep-th/0501025]

\bibitem{S2}
K. Kleidis, V. K. Oikonomou, 
Astrophys Space Sci (2016) 361:326 DOI 10.1007/s10509-016-2914-x, [arXiv:1609.00848 [gr-qc]]

\bibitem{S2_1}
S. D. Odintsov, V. K. Oikonomou, Class. Quant. Grav. 33, 125029, (2016), DOI:10.1088/0264-9381/33/12/125029 [arXiv: 1602.03309]; S. D. Odintsov, V. K. Oikonomou, Phys. Rev. D 92,124024 (2015), DOI:10.1103/PhysRevD.92.124024 [arXiv: 1510.04333]; S. D. Odintsov, V. K. Oikonomou, Phys. Rev. D 92, 024016 (2015), DOI:10.1103/PhysRevD.92.024016 [arXiv: 1504.0686]; K. Bamba et al, JCAP 0810, 045 (2008), DOI:10.1088/1475-7516/2008/10/045 [arXiv: 0807.2575].

\bibitem{S3}
J.B. Jimenez et al, Eur. Phys. J. C 76:631 (2016), DOI 10.1140/epjc/s10052-016-4470-5.

\bibitem{GP}
A. Shafiello et al, Phys. Rev. D 85, 123530 (2012); T. Yang et al, Phys. Rev. D 91, 123532 (2015).

\bibitem{S4}
M.P. Dabrowski and T. Denkieiwcz, Phys. Rev. D 79, 063521 (2009); M. P. Dabrowski and T. Denkiewicz, AIP Conf. Proc. 1241, 561 (2010); L. Fernandez-Jambrina, Phys. Lett. B 656, 9 (2007).

\bibitem{S5}
J.B. Jimenez et al, Cosmological future singularities in interacting dark energy models, arXiv:1607.06389 [gr-qc].

\bibitem{Om}
X. Zheng et al, Astrophys. J. 825, 17 (2016).

\end{thebibliography}
\end{document}